\documentclass[
amsmath,amssymb,
aps, physrev,
pra,
floatfix,
]{revtex4-2}
\usepackage{lipsum, babel}
\usepackage[utf8]{inputenc}
\usepackage{graphicx}
\usepackage{mathrsfs}
\usepackage{placeins}
\usepackage{rsfso}
\usepackage{hyperref}
\usepackage{amsmath}
\usepackage{subcaption}
\usepackage{dcolumn}
\usepackage{bm}
\usepackage{xcolor}
\usepackage{float}   
\begin{document}

\title{\textbf{Revealing Hidden Topology of Complex Vector Beams via Plasmonic Interactions} 
}%

\author{
Sahil Sahoo$^{1}$,
Ahmed Lafeef Ettapuram Naduvilepurayil$^{1}$,
Andre Yaroshevsky$^{1}$,
Peter Banzer$^{2}$,
Yuri Gorodetski$^{1,3,*}$
}

\affiliation{$^{1}$Department of Electrical and Electronic Engineering, Ariel University, Ariel 40700, Israel,}
\affiliation{$^{2}$Institute of Physics, University of Graz, NAWI Graz, Graz, Austria}
\affiliation{$^{3}$Department of Mechanical Engineering and Mechatronics, Ariel University, Ariel 40700, Israel.}

             \email{Contact author: yurig@ariel.ac.il}
\begin{abstract}

Structured light beams with space-variant polarization can be efficiently generated using voltage-tunable nematic liquid-crystal ($Q$-plate). By appropriately selecting the input state and the retardation of the $Q$-plate, an optical field acquires a spatially structured polarization distribution that is capable of encoding non-trivial topological information across the beam profile. These features can be directly read out through interaction with plasmonic nano-structures, such as circular and spiral slits. Here we show that, upon illumination, polarization-dependent excitation of surface plasmons converts the hidden topology of the polarization structure into observable intensity distributions, including plasmonic vortices and characteristic interference patterns, while the tunability of the input parameters enables a rich variety of distinct topological forms.

\end{abstract}
\maketitle

\section{Introduction}
The ability to engineer both the phase and the amplitude of light has driven a rapid progress in the field of structured light in recent years. Among various degrees of freedom available for shaping of electromagnetic fields, polarization holds particular importance \cite{forbes2021structured,he2022towards,chekhova2021polarization,shen2019optical}. Spatial variations of the polarization state the beam’s cross-section give rise to complex field distributions such as vector vortex beam \cite{rubinsztein2016roadmap,naidoo2016controlled,corona2024generation}. These beams combine both polarization and phase structures, thereby extending the versatility of structured light. Complex fields have become fundamental ingredients to a wide range of applications, including optical trapping and manipulation \cite{alpmann2015elegant, liesener2000multi, dufresne1998optical}, quantum information processing \cite{mair2001entanglement, mirhosseini2015high, harter2017parity, li2021liquid}, and advanced imaging methodologies \cite{milione20154, yang2017complex}. Vector beams (VBs) are structured optical fields in which the state of polarization varies spatially across the transverse plane. In general, they can be formed through a coherent superposition of distinct optical modes possessing orthogonal polarization states \cite{yao2025generation,yi2015hybrid,cardano2012polarization,han2024controllable,fickler2024higher}. This superposition leads to a non-separable coupling between the spatial phase and polarization degrees of freedom. Such fields can be regarded as a special class of Poincar\'{e} beams, characterized by spatially varying polarization states mapped onto the Poincar\'{e} sphere. The generation of such VBs can be achieved using $Q$-plates composed of nematic liquid crystals (LCs), which act as voltage-tunable retarders, enabling precise control of the output polarization \cite{bauer2015observation, nivas2017surface, marrucci2006optical, quiceno2020analysis, yao2025generation, shu2017polarization}.

VBs can be characterized using interferometric techniques, nonlinear and anisotropic \cite{quinto2023interferometric,al2022single,camacho2016nonlinear,aita2025longitudinal} light–matter interactions, as well as through coupling to metallic sub-wavelength structures, where the incident field excites collective surface-confined electron oscillations known as surface plasmons (SPs) \cite{ebbesen1998extraordinary,prinz2023orbital,murray2007plasmonic}. 
For instance illuminating a metallic circular slit by a VBs with azimuthal or radial polarization produces a dark or bright central spot corresponding to different Bessel modes of zero or first order respectively. Bessel SP distributions, often called plasmonic vortices (PV) due to their spiral phasefront of the form $\exp(il\varphi)$, are classified by their topological charge, $l$, and commonly known for carrying an angular momentum (AM) of $l\hbar$ per photon. 
It has been recently shown that plasmonic vortices can be generated by a circular slit structure illuminated by a circularly polarized light owing to the excitation of a polarization-dependent Pancharatnam-Berry (PB) phase \cite{gorodetski2008observation,prinz2023orbital,gorodetski2013generating}. Locally, the circular slit can be thought of as a spatially rotating polarizer with angular dependence of $\theta(\varphi) = \pi/2+\varphi$ resulting in a spiral PB phase of $\phi = \sigma \varphi$, where $\sigma = \pm 1$ represent the spin of the incident light (right circular polarization (-1) and left circular polarization (+1)). 
In the case of azimuthal (radial) polarization the electric field vector is always parallel (perpendicular) to the slit therefore producing no PB phase. 
This special form of the VB is represented by a trajectory lying on the equator of the Poincar\'{e} sphere \cite{lerman2009demonstration,yin2005subwavelength,chen2009plasmonic}. In a general case, VBs encompass a variety of polarization states, forming closed loops of arbitrary shape on the surface of the Poincar\'{e} sphere. As a result, they exhibit spatially varying structure with different polarization angle, handedness and ellipticity across the beam cross-section, therefore we refer them as complex vector beams (CVBs). This structured nature imparts CVBs with distinct topological properties.

In this Letter, we investigate the topological properties of CVB generated in free space by a $Q$-plate. The topology is an intrinsic property of the CVB, while the plasmonic slit structures are employed solely as sensitive analyzers that read out the presence of trivial/non-trivial topology. We show that, depending on the input parameters, the resulting CVB can carry a well-defined non-trivial topology, which can be detected upon illumination of spiral or circular plasmonic structures. Specifically, tuning the $Q$-plate retardation via the applied voltage leads to the appearance or disappearance of plasmonic vortices, indicating a topological phase change of the incident beam. The topological nature of the effect is further verified by exciting a spiral slit, which introduces an additional dynamic phase. Further tunability of the CVB is achieved by varying the incident polarization state or by introducing an external retardation element. As a result, we demonstrate a variety of CVBs whose polarization states trace\rotatebox[origin=c]{45}{$\infty$}- shaped loops and reveal their associated topological properties.
 
To summarize, the use of voltage-tunable $Q$-plate with standard polarizing elements enables generation of topologically encoded CVBs that can be probed via interaction with plasmonic nanostructure \cite{spektor2021orbital,yang2024plasmonic,vanacore2019ultrafast}. Beyond its fundamental importance, this approach holds great promise for applications in quantum information, nanoscale optical manipulation, and the development of advanced photonic devices that exploit topological phase engineering \cite{wang2020generating,zhou2023simultaneous,zhang2017radially,zhang2019nanoscale,zhou2024optical,abdollahramezani2020meta,jisha2021geometric}. 

\section{Generation and Probing of Topologically Encoded CVBs}
\begin{figure}[ht!]
\centering
\includegraphics[width=0.7\columnwidth]{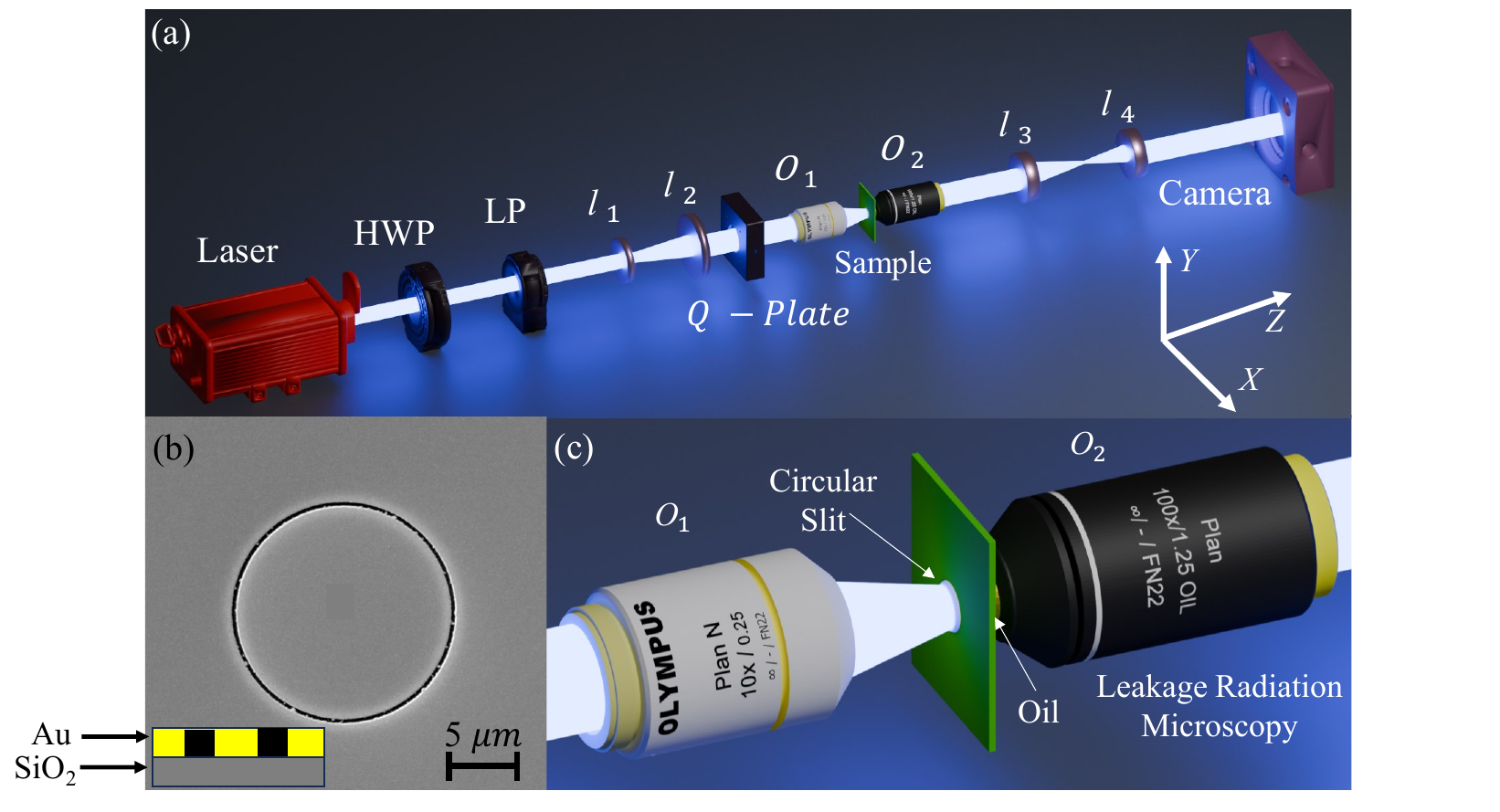}
\caption{\label{Experiment} 
Experimental setup: (a) Scheme of the experimental arrangement showing the illumination path of the laser beam through the $Q$-plate onto the sample. (b) SEM image of the fabricated circular slit sample. (c) Leakage radiation microscopy (LRM) configuration used to record the SPs field distribution.}
\end{figure}

The CVBs used in our experiment was generated as follows. A spatially filtered and collimated laser beam at a wavelength of $\lambda_0 = 780$ nm was passed through a half-wave plate (HWP) followed by a linear polarizer (LP) to prepare the desired input polarization state. The beam was then incident on a $Q$-plate with a topological charge of $q = 0.5$ (see Fig. \ref{Experiment}(a)). Two lenses, $l_1$ and $l_2$, were used to expand the beam in order to match the aperture size of the $Q$-plate. To probe the CVB topological phase, we used a 15 $\mu$m diameter plasmonic circular slit in order to excite the SP field distribution (Other slit diameters have been also tested however this parameter did not affect the results). The slit was fabricated using focused ion beam (FIB) milling on a 60~nm-thick gold film deposited on a glass substrate, as shown in the SEM image in Fig. \ref{Experiment}(b).

The CVB was initially focused to match the spot size to the structure diameter by using objective ($O_1$) with numerical aperture (NA) $0.25$. Then, the interference pattern of the generated SPs in the center of the circular slit was recorded from the glass side using leakage radiation microscopy (LRM) with an oil immersion objective $O_2$ with NA = $1.25/ \times 100$ magnification  \cite{hohenau2011surface}, as illustrated in Fig. \ref{Experiment}(c), and $l_3$ and $l_4$ were used as a Rayleigh imaging system to relay the image onto the camera. The retardation of the $Q$-plate was tuned by varying the applied voltage, enabling control of the spatial polarization distribution. The $Q$-plate's main axis was aligned along the $x$-axis for technical reasons. In theory, when the retardation of $\pi$ is provided by the device, incoming horizontal polarization ($\lvert H \rangle$) is expected to be fully converted to a radially polarized vector beam. Such distribution of the electric field presumably excites phaseless plasmonic wavefront with a typical constructive interference in the center.
Experimentally, three distinct input-voltage regimes are observed, each producing a different plasmonic interference field.

\begin{figure}[ht!]
\centering
\includegraphics[width=0.6\linewidth]{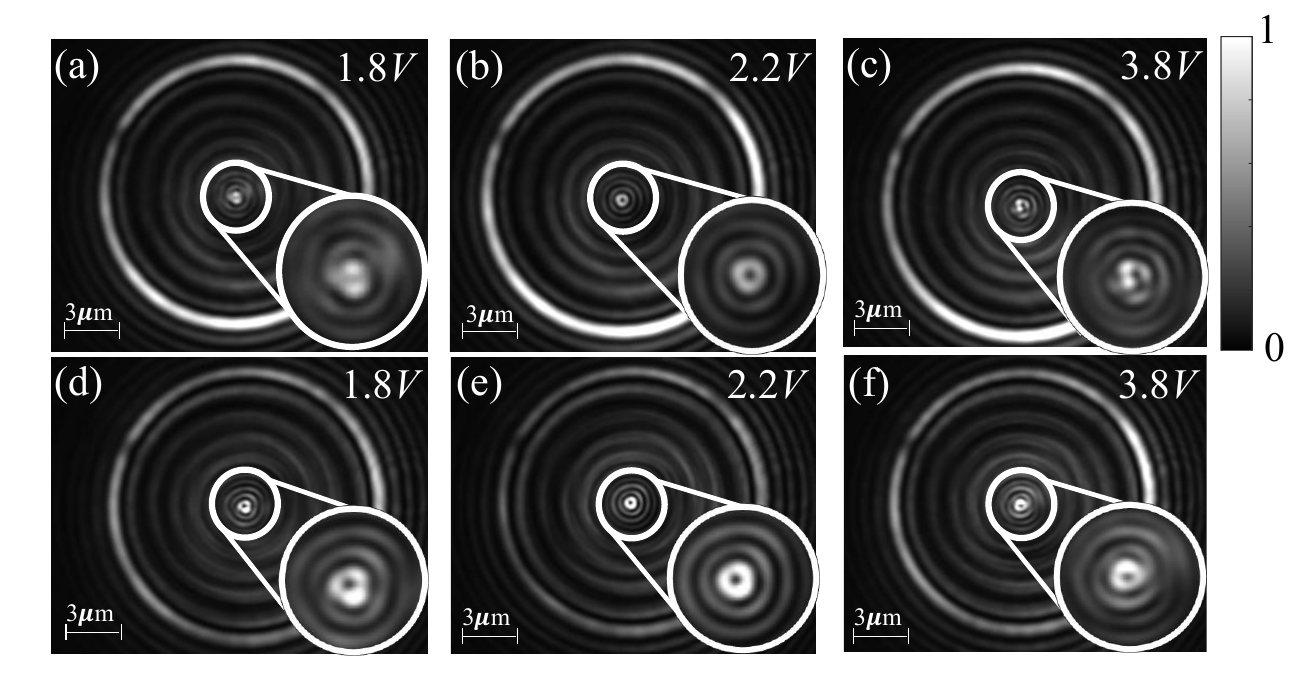}
\caption{\label{Plasmonic Interfernce}
Surface plasmon interference: (a–c) Experimentally measured SPs field for $|H\rangle$ and (d-f) for $|V\rangle$ distributions at applied voltages of 1.8$V$, 2.2$V$, and 3.8$V$, respectively.}
\end{figure}

When a voltage of $V_Q = 1.8V$ is applied, we indeed observe phaseless interference, manifested by a pronounced bright spot at the center (corresponding to a PV with a topological charge $l = 0$). Increasing this value to $V_Q = 2.2V$ results in the formation of a first-order PV, characterized by a central phase singularity (indicating a PV with a topological charge $l = \pm 1$). In order to verify the exact magnitude of the topological charge an additional dynamic phase induced by a spiral slit is used in the following section. Upon increasing the voltage further to $V_Q = 3.8V$, a phaseless plasmonic distribution reappears. The corresponding experimental results are shown in Figs. \ref{Plasmonic Interfernce}(a–c), with the insets highlighting the zoomed central region of each mode. In contrast, for vertical polarization ($|V\rangle$) incident on the $Q$-plate, the plasmonic field consistently exhibits a destructive plasmonic interference at the center for all applied voltages, as shown in Figs. \ref{Plasmonic Interfernce}(d–f). This latter behaviour is theoretically expected for an azimuthal polarization distribution due to the out-of-phase fields exciting SPs at the opposite sides of the circular slit \cite{lerman2009demonstration,chen2009plasmonic}. 

\begin{figure*}[ht!]
\centering
\includegraphics[width=0.9\linewidth]{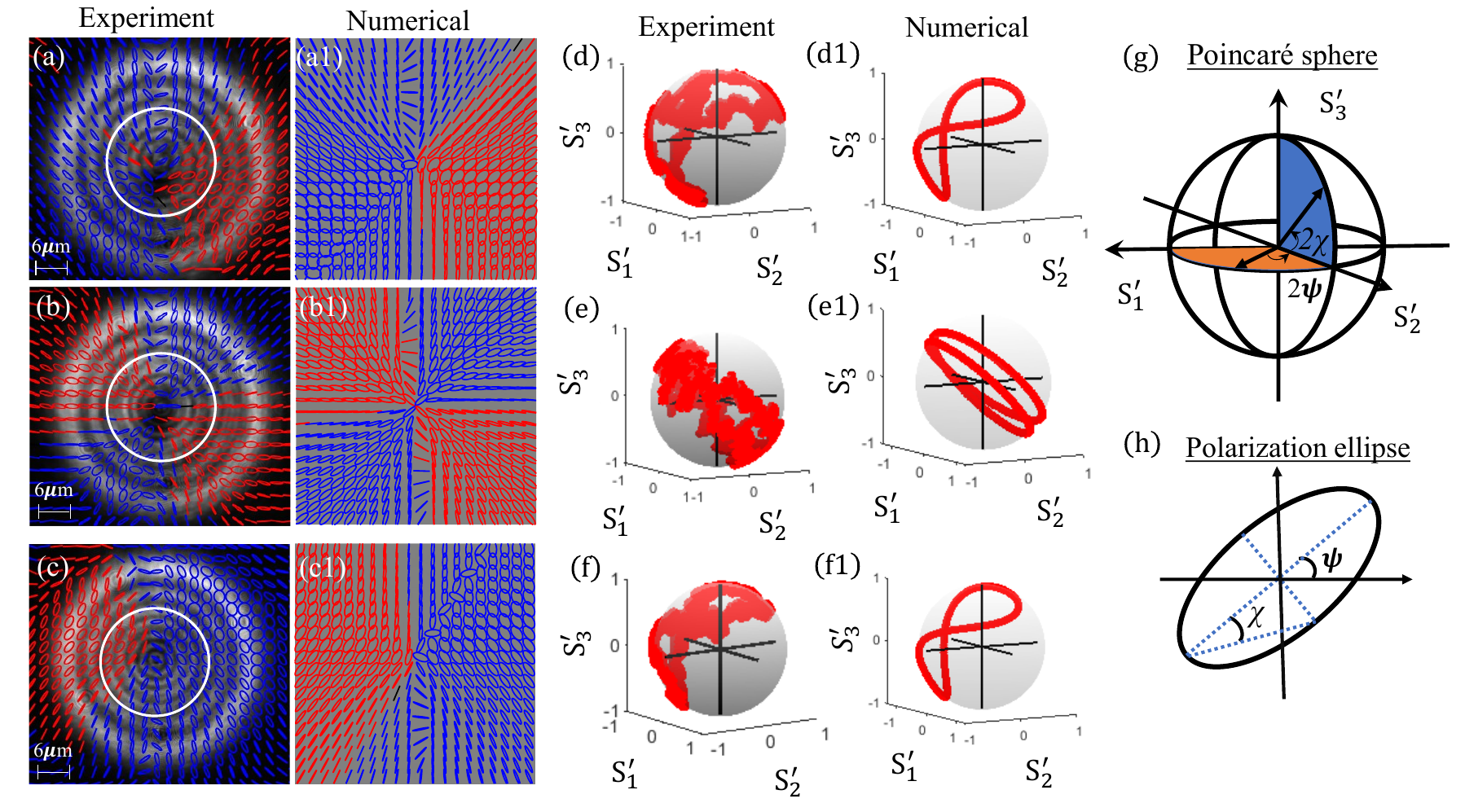}
\caption{\label{fig:poincaresphere} Experimental and numerical Stokes parameters: (a–c) Experimental measurements at applied voltages of 1.8$V$, 2.2$V$, and 3.8$V$, respectively, shown together with numerical results for phase retardations of $\delta \approx 1.5\pi, \pi$ and $0.5\pi$. (d–f) Measured Stokes parameters mapped onto the Poincar\'{e} sphere for three voltages and (d1-f1) calculated paths for the respective $\delta$ values. (g) $\psi$ and $\chi$ representation on the Poincar\'{e} sphere. (h) Polarization ellipse formation by $\psi$ and $\chi$.}
\end{figure*}

To elucidate the topological origin of the measured phenomenon we investigate the evolution of the Stokes parameters in free space across the beam after passing $Q$-plate with three values of the applied voltage. We use a fundamental four measurements technique to derive the Stokes' vector -- $[S_0, S_1, S_2, S_3]$ and then normalize it ($S_i' = S_i/S_0 $) by considering the light to be fully polarized \cite{stokes1851composition}. The polarization ellipse is then calculated from the obtained Stokes parameters distributions for each case. Figs. \ref{fig:poincaresphere}(a-c) show the experimentally measured polarization ellipses overlaid on the corresponding intensity maps for the $|H\rangle$ input. These measurements were performed with the sample removed from the setup in order to directly observe the free-space polarization structure of the CVBs. The red/blue color was chosen to represent the right/left helicity handedness while the ellipse aspect ratio is interpreted as the ellipticity. 
We note that the most significant polarization evolution is observed along the azimuthal direction ($\theta$), therefore we pick up points at a constant radius from the beam center ($r_0 = 11\mu m$, indicated by the white circle in the experimental results) and map them onto the Poincar\'{e}   sphere.  
As can be seen in Figs. \ref{fig:poincaresphere}(d-f) for each voltage we obtain a closed path characterizing the CVB topology.  The correspondence between the Poincar\'{e}  path and the cross section polarization map can be seen when recalling that the ellipse orientation corresponds to the azimuthal angle $2\psi$ while the helicity handedness is represented by the latitude angle $2\chi$. 

In attempt to compare three specific cases obtained with different voltages of $V_Q = 1.8V$, $2.2V$ and $3.8V$ the Poincar\'{e}-sphere trajectories of the generated CVBs are carefully examined. We note that in cases where a phaseless spot appears, the Stokes parameters trace a characteristic\rotatebox[origin=c]{45}{$\infty$}-shaped trajectory on the sphere. In contrast, at $V_Q = 2.2V$ — corresponding to a dark central region — the trajectory expands and nearly closes into a circular double loop. Importantly, this path does not lie exactly on the equatorial plane (as expected from a pure radial state) but is significantly inclined, indicating a hybrid polarization structure. We believe that this tilt—most likely caused by parasitic retardation in the $Q$-plate—is a key factor governing the topological behavior of the CVB. Unlike ideal radial or azimuthal polarization states, which lie strictly on the equator of the Poincar\'{e} sphere and involve only linear polarization components, this tilt drives the polarization trajectory away from the equator and introduces ellipticity. As a result, the polarization evolution spans both hemispheres of the Poincar\'{e} sphere, indicating that variations in linear orientation are now accompanied by changes in ellipticity across the CVB. 

For $|V\rangle$ input polarization the path is qualitatively similar, but, as expected, it is lying on the opposite side of the sphere (see Appendix 1). As a result, the polarization ellipse major axis is parallel to the circular slit, which leads to a destructive interference at the center and consequently producing a persistent dark spot. For the numerical analysis the $Q$-plate is modeled as a stack of $N$ LCs layers, where each layer is rotated by an angle $\tfrac{\Theta}{N}$. At the final LCs layer, the cumulative rotation reaches $\Theta \approx 0.5\pi$ (here ``$\approx$'' denotes an uncertainty of $\pm 0.03\pi$ to match the experimental results). The voltage-dependent phase retardation is given as, $\delta(V) = \frac{2\pi}{\lambda}\,(n_e(V) - n_o)\,d$, where $d$ is the LCs thickness, $n_o$ the ordinary refractive index, and $n_e(V)$ the extraordinary index, which varies with the applied voltage. While the total retardation of $\delta$ is distributed across the whole stack, each layer contributes a retardation of $\tfrac{\delta}{N}$.  

The Jones matrix of one $Q$-plate cell is obtained by sequentially applying a rotation matrix $R$ to the elementary retardation matrix $J_0$:

\begin{equation}
J_{\Theta,\delta} = \prod_{n=1}^{N} R\!\left(-\tfrac{n\Theta}{N}\right) J_0\!\left(\tfrac{\delta}{N}\right) R\!\left(\tfrac{n\Theta}{N}\right),
\end{equation}

with  

\begin{equation}
R(\Theta) = \begin{bmatrix}
\cos\Theta & -\sin\Theta \\
\sin\Theta & \cos\Theta
\end{bmatrix}, 
\qquad
J_0(\delta) = \begin{bmatrix}
e^{i\delta/2} & 0 \\
0 & e^{-i\delta/2}
\end{bmatrix}.
\end{equation}

In the limit where $N$ is very large and $\delta \gg \Theta$, the $Q$-plate behaves as an effective uniform wave plate described by $M_{\Theta,\delta}$. The matrix can be expressed as $M_{\Theta,\delta} = R(-\Theta)\, J_0$, corresponding to a rotation to the final layer at angle $\Theta$, with the fast axis of $J_0$ aligned horizontally. In this regime, the retardation $\delta$ is defined modulo $2\pi$, as only its value within the interval $[0, 2\pi)$ influences the polarization state, while any additional integer multiple of $2\pi$ produces an equivalent transformation (e.g., $10001\pi \equiv \pi$) \cite{sit2024spatially}. To account for the spatial dependence, the Jones matrix is expressed as $J_f = R(-\varphi)\, M_{\Theta,\delta}\, R(+\varphi),$ where $\varphi = q \cdot \tan^{-1}\left(\frac{y}{x}\right) + \alpha,$ with $q$ the topological charge of the $Q$-plate and $\alpha$ the LC director orientation.
 
The aforementioned global tilt can be modeled by an additional retardation introduced after the $Q$-plate. 
We incorporate this feature into the numerical model by introducing a retarder $T_r$, with retardation $\gamma$ and orientation angle $\beta$ \cite{lerman2010generation},

\begin{equation}
T_r(\gamma,\beta) =
\begin{bmatrix}
\hspace{6pt} e^{i\frac{\gamma}{2}} \cos^2\beta + e^{-i\frac{\gamma}{2}} \sin^2\beta \hspace{6pt} &
\hspace{12pt} \left( e^{i\frac{\gamma}{2}} - e^{-i\frac{\gamma}{2}} \right) \sin\beta \cos\beta \hspace{6pt} \\[12pt]
\hspace{6pt} \left( e^{i\frac{\gamma}{2}} - e^{-i\frac{\gamma}{2}} \right) \sin\beta \cos\beta \hspace{6pt} &
\hspace{12pt} e^{i\frac{\gamma}{2}} \sin^2\beta + e^{-i\frac{\gamma}{2}} \cos^2\beta \hspace{6pt}
\end{bmatrix}
\end{equation}

\begin{figure}[ht!]
\centering
\includegraphics[width=0.7\linewidth]{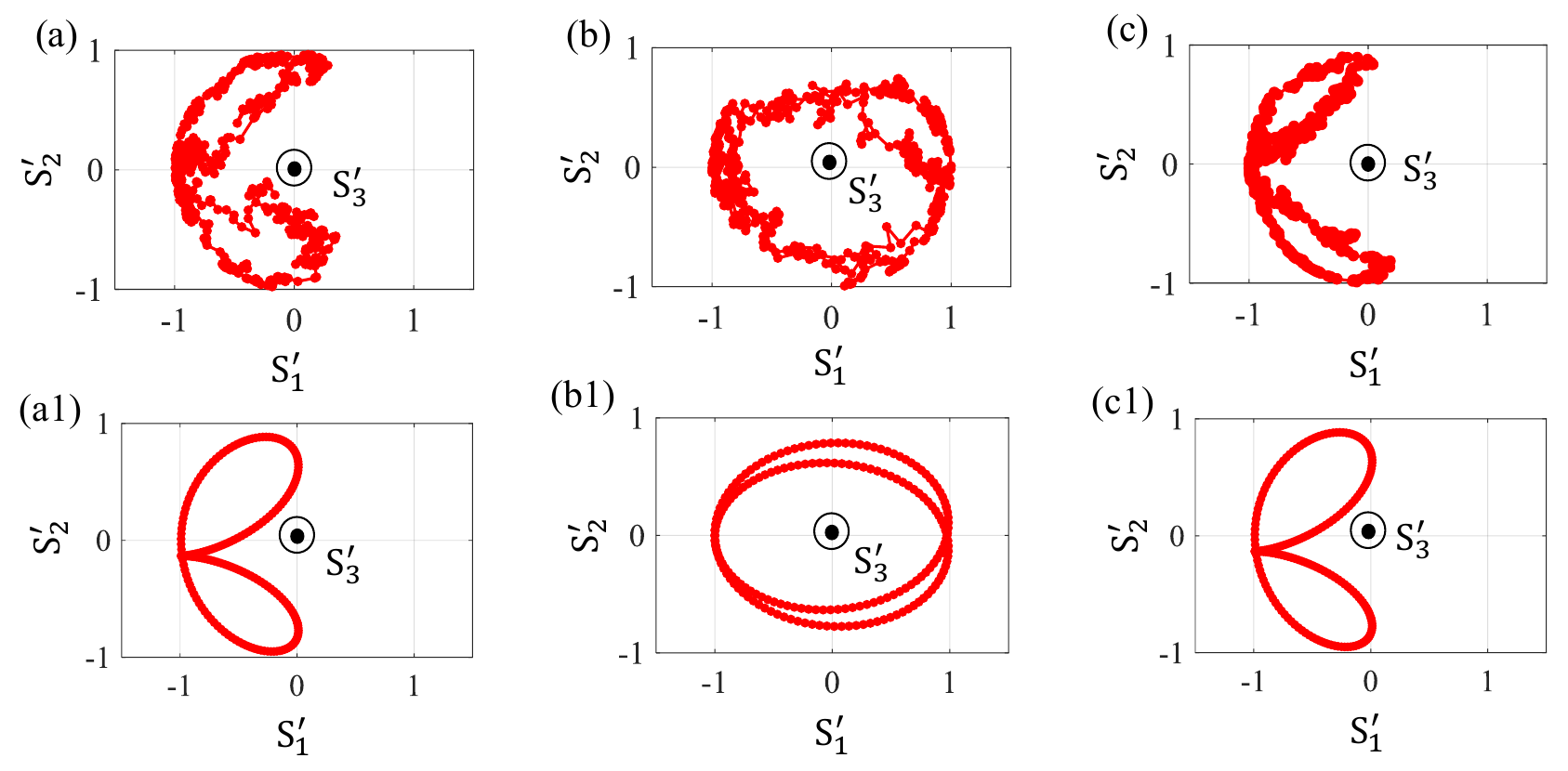}
\caption{\label{Encircling S_3} Encircling $S'_3$: Stokes parameters plotted on the $S'_2$–$S'_1$ plane through the $S'_3$ axis. Experimental results are shown for applied voltages $V_Q = 1.8V$, $2.2V$, and $3.8V$, together with numerical results corresponding to retardation $\delta \approx 1.5\pi$, $\pi$, and $0.5\pi$, respectively. }
\end{figure}

Using our experimental parameters $q=0.5$ and $\alpha = -\tfrac{\pi}{4},$ (to represent the $x$ - orientation of the plate) we find out that a retarder $T_r\left(\tfrac{3\pi}{4},\,0\right)$ should be used to properly represent the measured results.
Notably, the numerical model shows excellent agreement with the experimental results. For applied voltages of $1.8V$, $2.2V$, and $3.8V$ with a $|H\rangle$-polarized input, the measured Stokes parameters closely follow the theoretical predictions for the retardations $\delta \approx 1.5\pi$, $\pi$, and $0.5\pi$, respectively ("$\approx$" denotes an uncertainty of $\pm0.05\pi$). 
The corresponding results for a $|V\rangle$-polarized input are presented in Appendix A1. This comparison is shown in Figs. \ref{fig:poincaresphere}(a1–c1). When mapped onto the Poincar\'{e} sphere, the Stokes-parameter trajectories form a characteristic figure-“\rotatebox[origin=c]{45}{$\infty$}” loop near one of the poles for $\delta \approx 1.5\pi$ and $0.5\pi$, as illustrated in Figs. \ref{fig:poincaresphere}(d1) and \ref{fig:poincaresphere}(f1). As the retardation approaches $\delta \approx \pi$, the trajectory expands and nearly bisects the sphere, producing a loop that winds twice around the Poincar\'{e} sphere, as shown in Fig. \ref{fig:poincaresphere}(e1). The observed deviations from the theoretical trajectories are primarily attributed to pixel-level noise in the Stokes measurements, as each polarization state is reconstructed from individual pixel intensities.

\begin{figure*}[t]
\centering
\includegraphics[width = 1\linewidth]{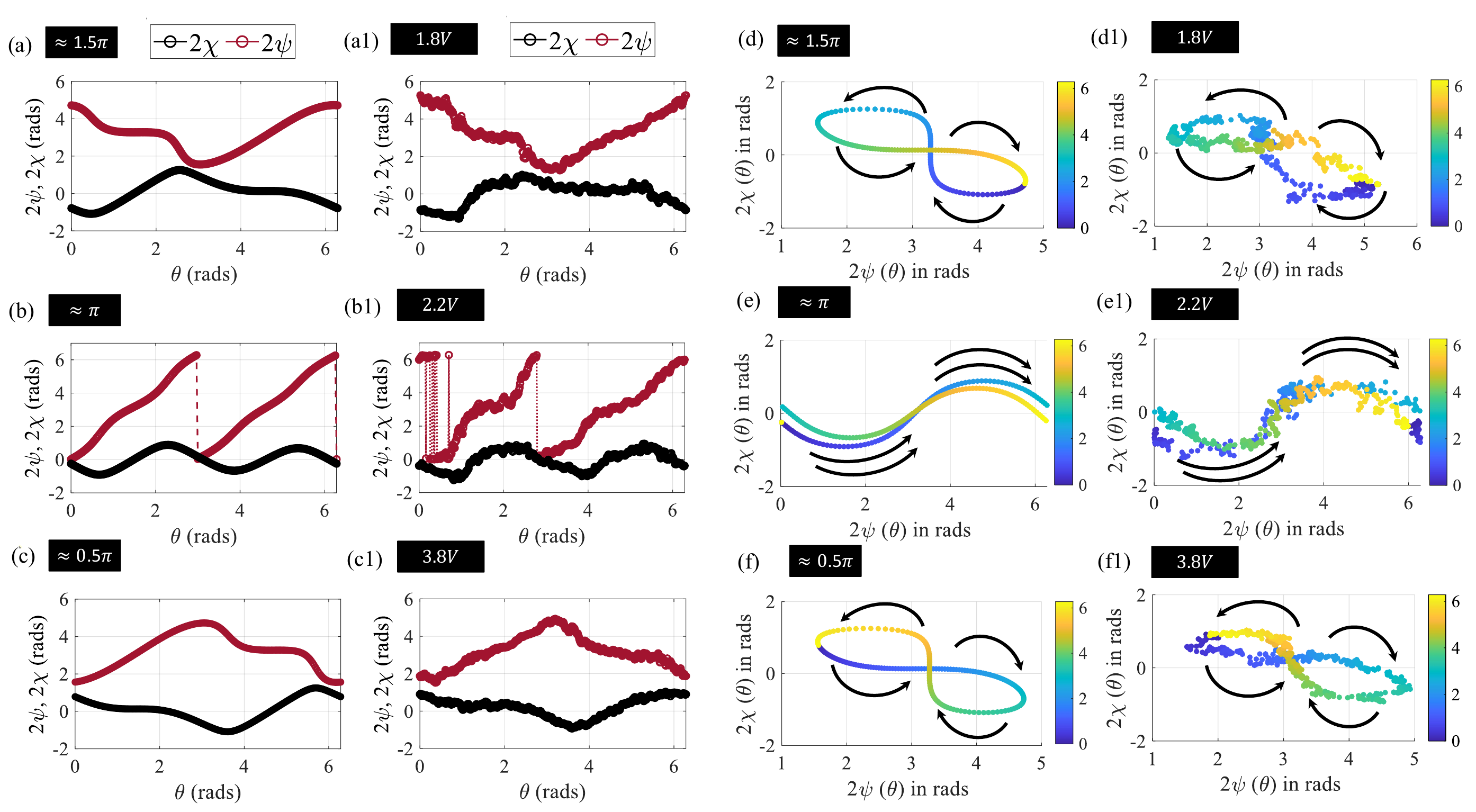}
\caption{\label{Polarization Winding} Stokes-vector winding: a–c Evolution of $2\chi$ and $2\psi$ with $\theta$ for $\delta \approx 1.5\pi$, $\pi$, and $0.5\pi$, respectively. Panels a1–c1 show the corresponding experimental results for applied voltages of $1.8V$, $2.2V$, and $3.8V$. d–f Numerical plots of $2\chi(\theta)$ versus $2\psi(\theta)$ for $\delta \approx 1.5\pi$, $\pi$, and $0.5\pi$, respectively, with the corresponding experimental results shown in d1–f1 for $1.8V$, $2.2V$, and $3.8V$.
}
\end{figure*}

A significant difference in geometry of the paths can be detected when inspecting the projection of the state points onto $S'_1-S'_2$ plane. ``Top view'' from the positive $S'_3$ axis reveals that only in the case of $V_Q = 2.2V$ the path fully encircles the axis twice while with other two voltages the closed\rotatebox[origin=c]{45}{$\infty$} - shape excludes it. Figs. \ref{Encircling S_3}(a-c) shows the projections for the $|H\rangle$ input polarization state (for $|V\rangle$-polarized input refer to A1).
This behavior is also confirmed numerically in Figs. \ref{Encircling S_3}(a1-c1). 
Mathematically speaking, the difference between the paths can be described by the behavior of the ellipse orientation angle.
Specifically, only when the angle $\psi$ evolves from $0$ to $2\pi$ radians an additional phase appears (at $V_Q = 2.2V$).
At this point we may suggest that the phase 
that has appeared in the plasmonic field can be linked to this topological feature of the path. 

A circular slit exciting SPs can be defined by the angle $\theta$ which defines a radial polarizer. Accordingly, when a $Q$-plate produces a perfect radial polarization no additional phase is expected. Nevertheless, our system does not provide such a state as can be seen from a slight elongation of the loop in Figs. \ref{Encircling S_3}(b,b1) caused by a path tilt. More importantly, the $2\chi$ angle (not represented in this top-view diagram) changes its sign twice. We suspect that a combined evolution of these two polarization angles may shed a light on a topological state of the CVB.   

Figs. \ref{Polarization Winding}(a-c) and (a1-c1) show the evolution of the angles $\psi$ and $\chi$ (we use $2\psi \equiv \tilde{2\psi} \pmod{2\pi}$) to identify ($-\pi$) and ($+\pi$) as the same polarization state, thereby avoiding an artificial jump when scanning continuously along the (x)-axis) as functions of the beam azimuthal angle $\theta$. For $V_Q = 1.8V$ with a $|H\rangle$-polarized input, corresponding to a retardation $\delta \approx 1.5\pi$, the Poincar\'{e}   trajectory occupies only a restricted portion of the equator of the Poincar\'{e}   sphere. In this regime, $2\chi$ appears effectively unbounded due to repeated handedness reversal, while $2\psi$ remains bounded, leading to a libration of the polarization vector around the equator. This combined behavior becomes evident when plotting $2\chi(\theta)$ versus $2\psi(\theta)$ ($\theta$ is represented by the color scale), which produces a characteristic\rotatebox[origin=c]{45}{$\infty$}-shaped loop, as observed in both numerical and experimental results shown in Figs. \ref{Polarization Winding}(d) and (d1). The presence of oppositely rotating lobes that do not complete a full winding indicates a trivial polarization topology of the beam.

In contrast, for $V_Q = 2.2V$, the Stokes parameters encircle the $S'_3$ axis twice. As a consequence, $2\psi$ evolves continuously over a full $2\pi$ range, while $2\chi$ exhibits a sinusoidal dependence on $\theta$ with an additional oscillation compared to the $V_Q = 1.8V$ case, as shown in Figs. \ref{Polarization Winding}(b) and (b1). The combined evolution, visualized by plotting $2\chi(\theta)$ versus $2\psi(\theta)$, reveals a continuous oscillatory trajectory along the equator, as shown in Figs. \ref{Polarization Winding}(e) and (e1). This behavior corresponds to a complete rotation of the polarization vector around the equator, indicating a non-trivial topology and resulting in a finite geometric phase imprinted on the beam. Interestingly, $2\chi$ spans both hemispheres, whereas it is otherwise bounded.

Finally, for $V_Q = 3.8V$, behavior analogous to that observed at $V_Q = 1.8V$ is recovered. As shown in Figs. \ref{Polarization Winding}(c) and (c1), the Stokes parameters again trace an\rotatebox[origin=c]{45}{$\infty$}-shaped loop, and the corresponding $2\chi(\theta)$ versus $2\psi(\theta)$ plots exhibit similar librational dynamics of the polarization vector across the beam, as illustrated in Figs. \ref{Polarization Winding}(f) and (f1).

In summary we observe an additional spiral phase arising in the plasmonic wavefront in the case 
when both $2\chi (\theta)$ and $2\psi(\theta)$ oscillate (like in the $\pi$ retardation case). As a result a PV of a topological charge $|l|=1$ is excited in the center of the circular slit. However, it is essential to experimentally validate the topology of this phase. Hereafter, we propose to analyze the discussed cases when considering an addition of a dynamic phase. 

\section{CVB with dynamic phase}
\begin{figure}[ht!]
\centering
\includegraphics[width= 0.6\linewidth]{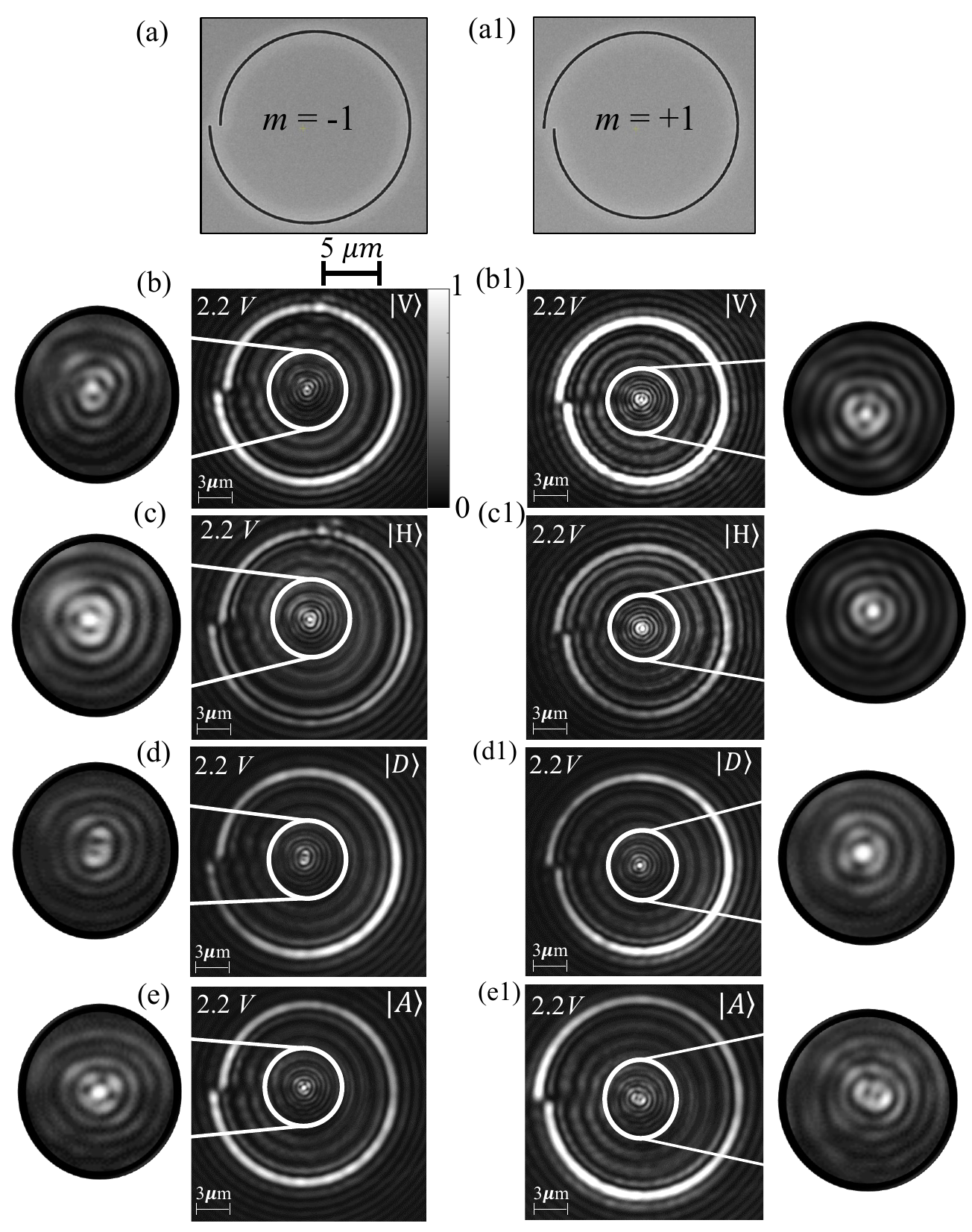}
\caption{\label{Dynamic Phase} Dynamic phase: (a,a1) SEM image of spiral $m = -1$ and $m = +1$ respectively. (b–e) SPs wavefront interference generated by a spiral structure with topological charge $m=-1$ for the input polarizations $|V\rangle$, $|H\rangle$, $|D\rangle$, and $|A\rangle$ at an applied voltage of $2.2V$. (b1–e1) Corresponding SP interference patterns for the opposite spiral handedness, $m=+1$, under the same polarization and voltage conditions.}
\end{figure}

We have fabricated a spiral nano-slit defined by angle dependent radius $r(\theta) = r + m\lambda_{SP}\theta/2\pi$, where $\lambda_{SP}$ is the plasmonic wavelength, $m$ is a spiral pitch and $r = 7.5 \mu m$ as shown in Figs .\ref{Dynamic Phase}(a,a1). The aim of such a structure is to induce an additional, polarization independent dynamic phase ramp in the excited field. Previously \cite{gorodetski2008observation} such a method had been used to detect a geometric phase resulted from an optical spin-Hall effect.
In our investigation, we use a slit with $m=+1$ and $m = -1$. In Fig. \ref{Dynamic Phase} we compare plasmonic wavefronts excited with CVBs generated by $|V\rangle$, $|H\rangle$, $|D\rangle$ (Diagonal polarization) and $|A\rangle$ (Anti-diagonal polarization) states with a $V_Q = 2.2$V. 
Interestingly, when using the $|V\rangle$ and $|H\rangle$ states the field distribution in the center seems to be indifferent to the spiral handedness. This effect can be explained by the fact that the plasmonic wave launched by the structure is not a pure Bessel mode but results from a superposition of vortices with various AM values. Nevertheless, when observing the measured intensities excited by the $|D\rangle$, $|A\rangle$ states one can clearly witness the topological charge summation rule - $l=l_{CVB} + m$ (here $l_{CVB}$ stands for the intrinsic topology of the CVB). In particular, we observe that by changing the spiral helicity or the CVB state, the plasmonic vortex change such that, for $m=-1$, the $l=-2$ to $l=0$ as the polarization change from $|D\rangle$ to $|A\rangle$. Conversely, for $m=+1$, $l$ change from $0$ to $+2$ for the same change from $|D\rangle$ to $|A\rangle$.

\begin{figure}[ht!]
\centering
\includegraphics[width= 0.7\linewidth]{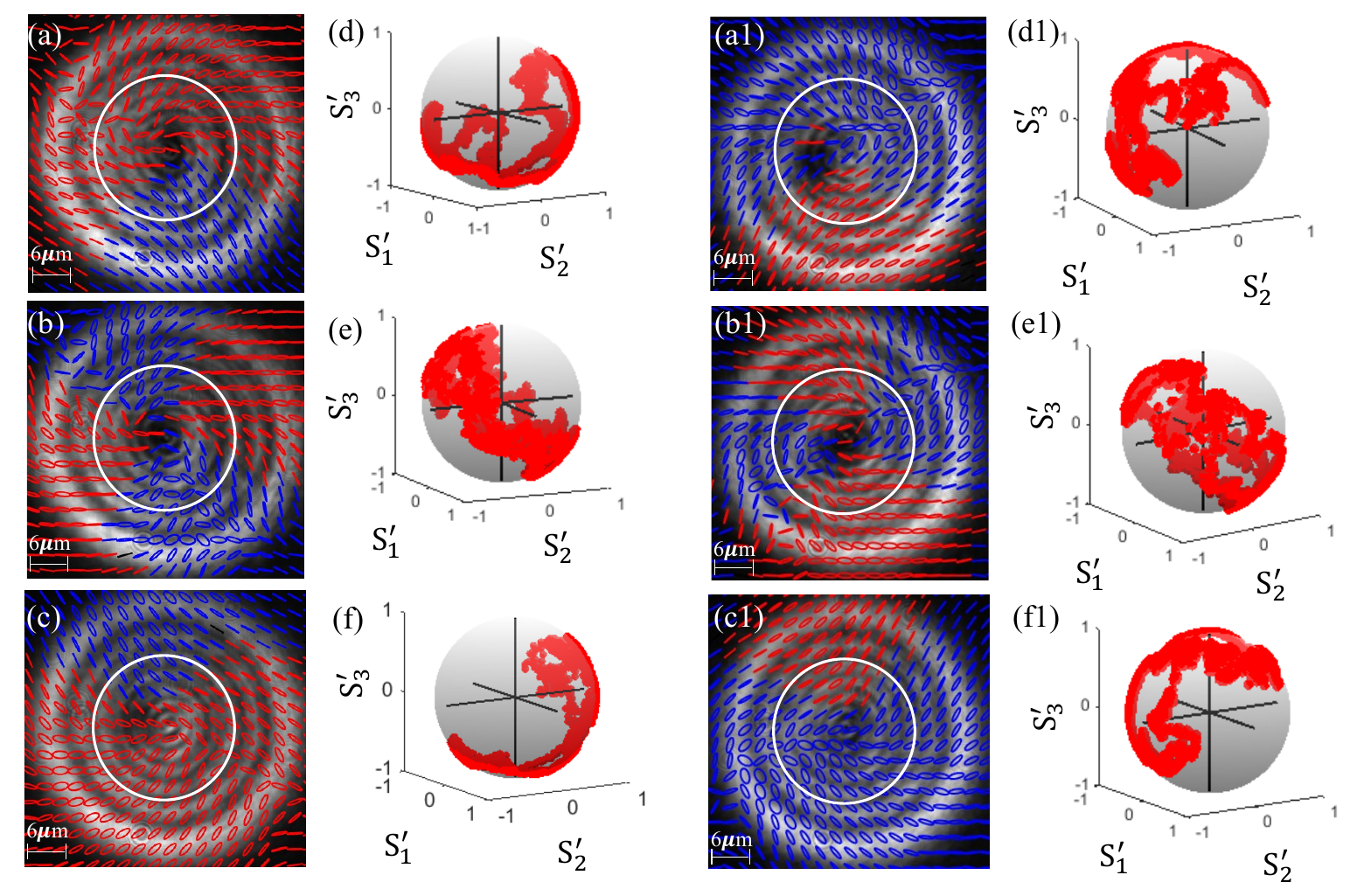}
\caption{\label{45,-45} $|D\rangle$ and $|A\rangle$ inputs: (a–c) Polarization maps for $|D\rangle$ at $1.8V$, $2.2V$, and $3.8V$. (d–f) Corresponding Stokes trajectories on the Poincar\'{e} sphere. (a1–c1) Polarization maps for $|A\rangle$ at  $1.8V$, $2.2V$, and $3.8V$. (d1–f1) corresponding Stokes trajectories.
}
\end{figure}

The effect of the retardation (provided by the $Q$-plate voltage) is presented in Fig .\ref{45,-45}. However, when inspecting the polarization maps in Figs .\ref{45,-45}(a-c;a1-c1) we note a significant chirality of the vector field, particularly in the central region of the beam. This field handedness, as we believe, lifts the degeneracy of the plasmonic vortex modes in Fig.\ref{Dynamic Phase}. The effect of the voltage is embodied in the ellipticity, so that for $V_Q = 2.2$V the helicity of the field changes sign two times while in the cases of $V_Q = 1.8,3.8$ V we observe positive $\chi$ values along one quarter of the beam cross-section. It is worth mentioning that since $|H/V\rangle = \frac{1}{\sqrt{2}}(|D\rangle \pm |A\rangle$ the result of the vector summation of the maps in Fig. \ref{45,-45} is expected to be exactly given by the maps in Fig. \ref{fig:poincaresphere}. However, while the vector field achieved with $|D/A\rangle$ inputs exhibits a well pronounced chirality, the fields induced by the $|H/V\rangle$ states are degenerated.

\begin{figure*}[ht!]
\centering
\includegraphics[width=0.9\linewidth]{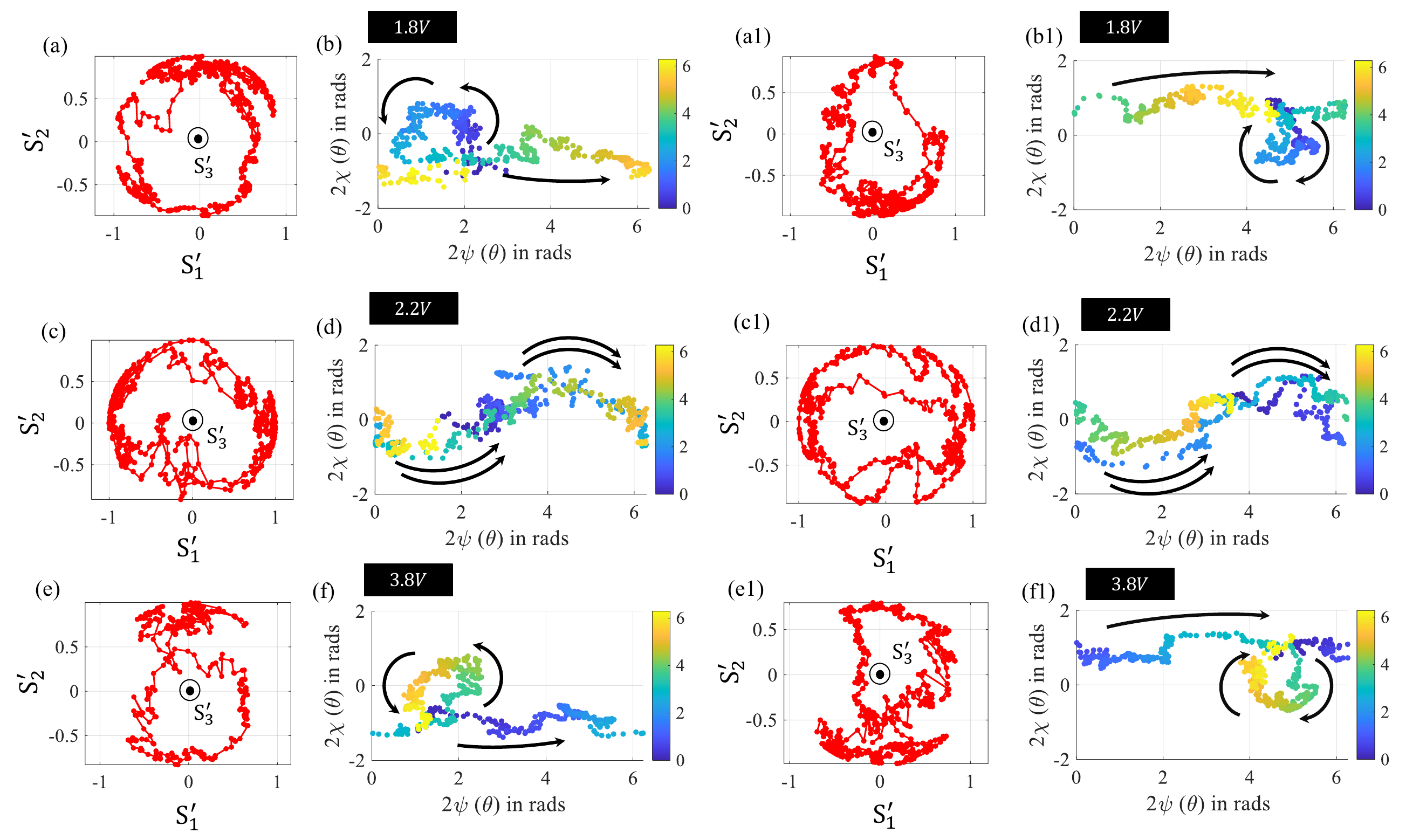}
\caption{\label{Encircled One Loop} Encircled one-loop polarization vectors:
Stokes-parameter trajectories on the $S'_2$–$S'_1$ plane and the corresponding $2\chi$–$2\psi$ plots for $|D\rangle$ and $|A\rangle$ inputs at $1.8V$ (a–b, a1–b1), $2.2 V$ (c–d, c1–d1), and $3.8V$ (e–f, e1–f1).}
\end{figure*}

The measured Poincar\'{e}-sphere trajectories for the corresponding CVBs are shown in Figs. \ref{45,-45}(d–f,d1–f1). For an applied voltage of 2.2 V, the trajectory again forms a double loop spanning a sphere diameter. In contrast, at 1.8 V and 3.8 V, the paths exhibit a \rotatebox[origin=c]{45}{$\infty$}-shaped structure in which only one of the loops encircles the $S_3$ axis.

This behavior can be clearly inferred from the top-view projections of the trajectories shown in Figs. \ref{Encircled One Loop}(a,c,e) for the $|D\rangle$ input. The corresponding $2\chi$ versus $2\psi$ plots in Figs. \ref{Encircled One Loop}(b,d,f) show that for $1.8V$ and $3.8V$ the trajectories form a single closed loop, whereas for $2.2V$ a sinusoidal dependence is obtained. This sinusoidal behavior reflects the rotation of polarization vectors observed at this voltage, corresponding to motion along the equatorial region of the Poincar\'{e} sphere. Interestingly, when only one of the loops is closed while the other remains open, the polarization vector evolution represents a mixed regime combining libration and rotation behavior. In this case, the loop that encircles the $S'_3$ axis remains confined to a single hemisphere, corresponding to a fixed sign of $\chi$. This hemispherical confinement becomes evident when the beam illuminates the spiral structure. A similar encircling behavior is observed for the $|A\rangle$ input, as shown in Figs. \ref{Encircled One Loop}(a1,c1,e1) and the corresponding $2\chi$–$2\psi$ plots in Figs. \ref{Encircled One Loop}(b1,d1,f1), with the main difference being a complementary sign of $\chi$.

\begin{figure}[ht!]
\centering
\includegraphics[width=0.6\linewidth]{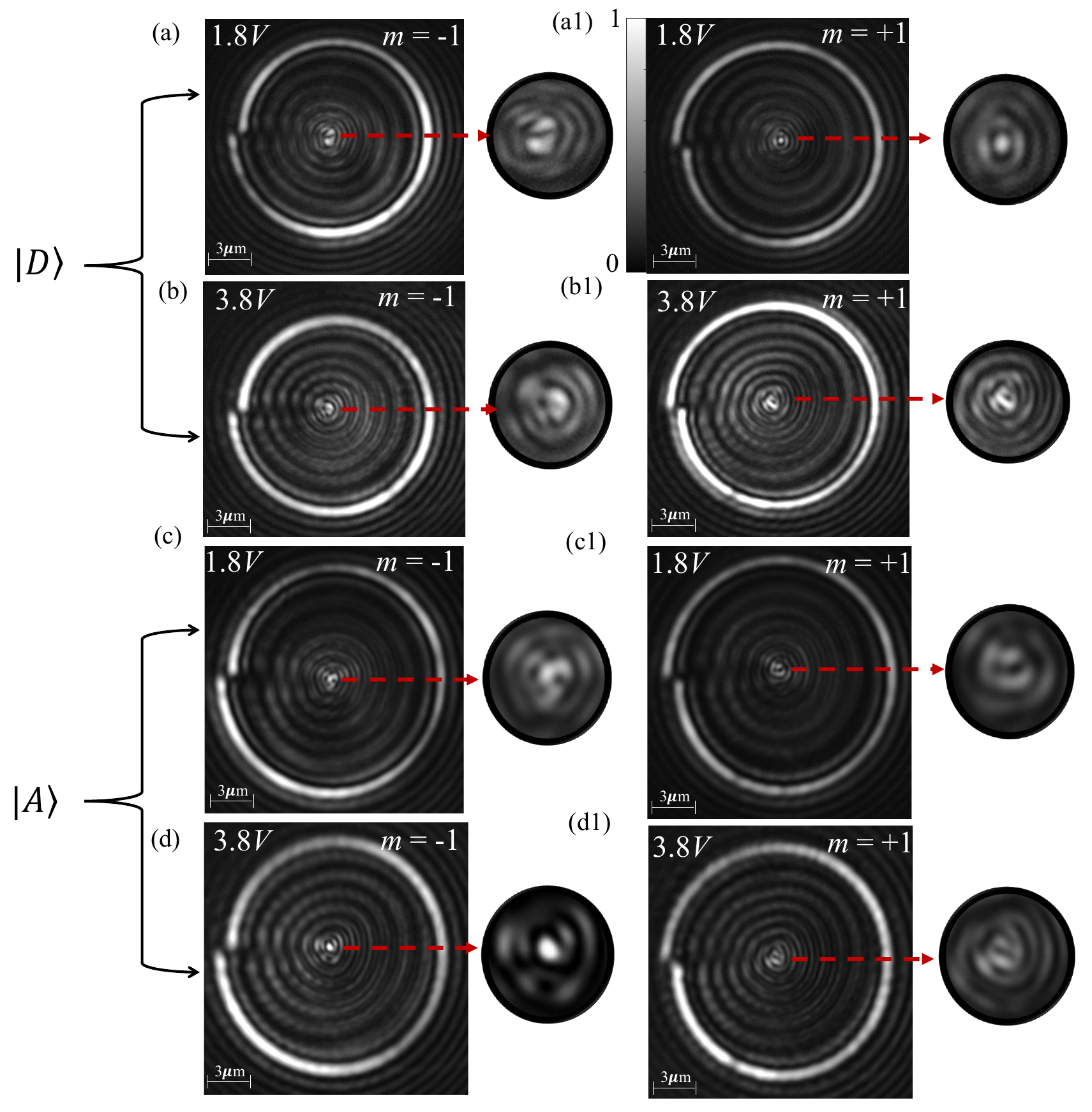}
\caption{\label{partial phase} Partial phase:
SP wavefront interference for $|D\rangle$ and  $|A\rangle$ inputs at 1.8 V and 3.8 V. Panels (a,b) and (c,d) correspond to $|D\rangle$ and $|A\rangle$ inputs, respectively, for a spiral with $m=-1$, while (a1,b1) and (c1,d1) show the corresponding results for the opposite handedness, $m=+1$.}
\end{figure}

When the spiral structure is illuminated at $1.8V$ and $3.8V$ with $|D\rangle$ and $|A\rangle$ inputs for both $m=-1$ and $m=+1$, a partial dislocation of the phase is observed, as shown in Fig. \ref{partial phase}. The presence of only one closed loop in the $2\chi$–$2\psi$ representation, with the remaining part of the trajectory remaining open, indicates a partial phase accumulation. This corresponds to an incomplete formation of states with $l=2$ or $l=0$. Furthermore, since the sign of $\chi$ for the $|D\rangle$ and $|A\rangle$ inputs is opposite, complementary phase dislocation patterns are observed in the corresponding cases.

\begin{figure}[ht!]
\centering
\includegraphics[width=0.7\linewidth]{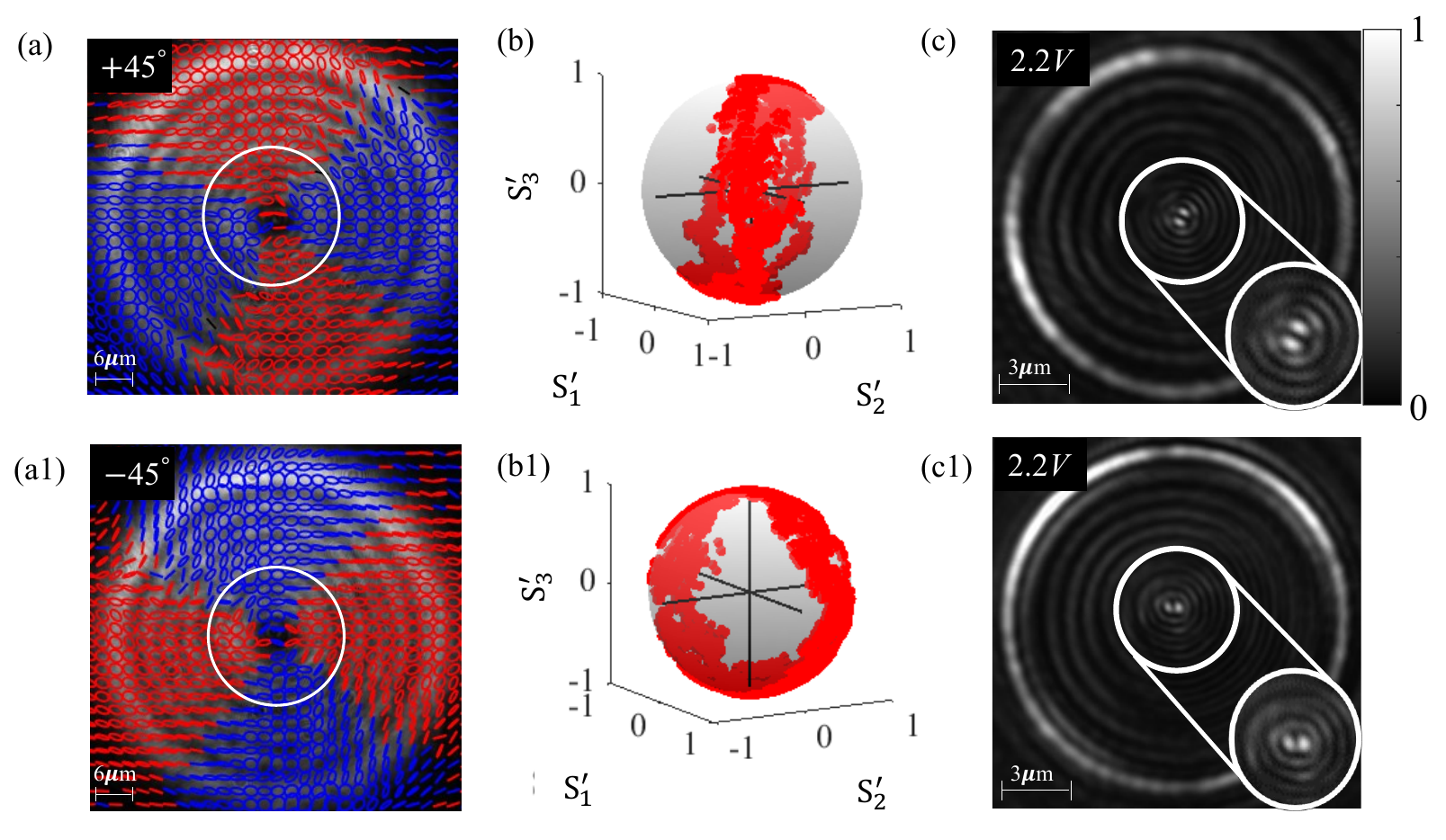}
\caption{\label{Tilting} Output-state tilting: (a,a1) Stokes parameters for a quarter-wave plate (QWP) oriented at $\pm45^\circ$. (b,b1) Corresponding Stokes trajectories on the fig:poincaresphere, oriented perpendicular to the equator. (c) Plasmonic field distributions at $2.2\,\mathrm{V}$ for QWP orientations of $\pm45^\circ$.
}
\end{figure}

As discussed earlier, for all the cases where a plasmonic vortex is observed the corresponding Stokes trajectories encircle the $S'_3$ axis twice. We wish to generate a CVB such that its polarization state does not encircle the $S'_3$ axis however completes a double loop at the Poincar\'{e} sphere diameter. To achieve this, the retardation is adjusted by placing a quarter-wave plate (QWP) after the $Q$-plate, which rotates the output polarization path to be orthogonal to the equator, thereby eliminating the encircling.

The resulting output ellipse parameters for $\pm45^\circ$ QWP orientations are shown in Fig. \ref{Tilting}(a,a1). The corresponding transformation is given by
\begin{equation}
\mathbf{J}
=
J_{\mathrm{QWP}}(\pm45^\circ)\cdot
T_r\!\left(\tfrac{3\pi}{4},\,0\right)\cdot
J_f .
\end{equation}
The associated trajectories on the fig:poincaresphere for the $|V\rangle$ input are shown in Figs. \ref{Tilting}(b, b1). The plasmonic field excited by this distribution has a clear dipolar shape like if it was generated by a linear polarization. This might be understood by inspecting the polarization maps. One can analyze this effect as a superposition of two modes resulted from a spin Hall effect with opposite circular input. The polarization maps demonstrate a clear dominance of circular polarizations. The result is then observed as a sum of two vortices with $l = \pm 1$ or if it was initially excited by a linear state \cite{lerman2009demonstration}.

\section{Conclusion}

We demonstrated the ability to encode topological information in a CVBs by tailoring the input polarization using a $Q$-plate. The beam phases were manipulated through several degrees of freedom, including the voltage-controlled retardation, the incident polarization state, and an additional waveplate. The resulting structured beam was probed by a plasmonic device in the form of a circular slit, which excited different plasmonic vortex modes. The achieved topologies of the CVBs were classified by the shape of the corresponding trajectories on the Poincar\'{e} sphere. Specifically, the topological charge of the plasmonic vortex was found to be directly related to how the polarization path circumvents the $S_3$ axis. Due to a parasitic global retardation in the $Q$-plate, the polarization trajectories are tilted away from the equatorial plane of the Poincar\'{e} sphere, preventing the realization of ideal radial and azimuthal polarization states. Instead, the generated beams exhibit a controlled ellipticity while preserving their underlying topological structure. We believe that the ability to engineer and read out such space-variant polarization states may be useful for a wide range of photonic applications, including information encoding and sensing.

\section{DATA AVAILABILITY}
The data supporting the findings of this study are not publicly available but can be obtained from the authors upon reasonable request.

\section{Acknowledgment}
Authors acknowledge the Israeli Ministry of Innovation, Science and Technology for funding the research

\newpage
\section{Appendix}
\section*{A1: Stokes parameters for $|V\rangle$ input}

\begin{figure}[ht!]
\centering
\includegraphics[width=0.6\linewidth]{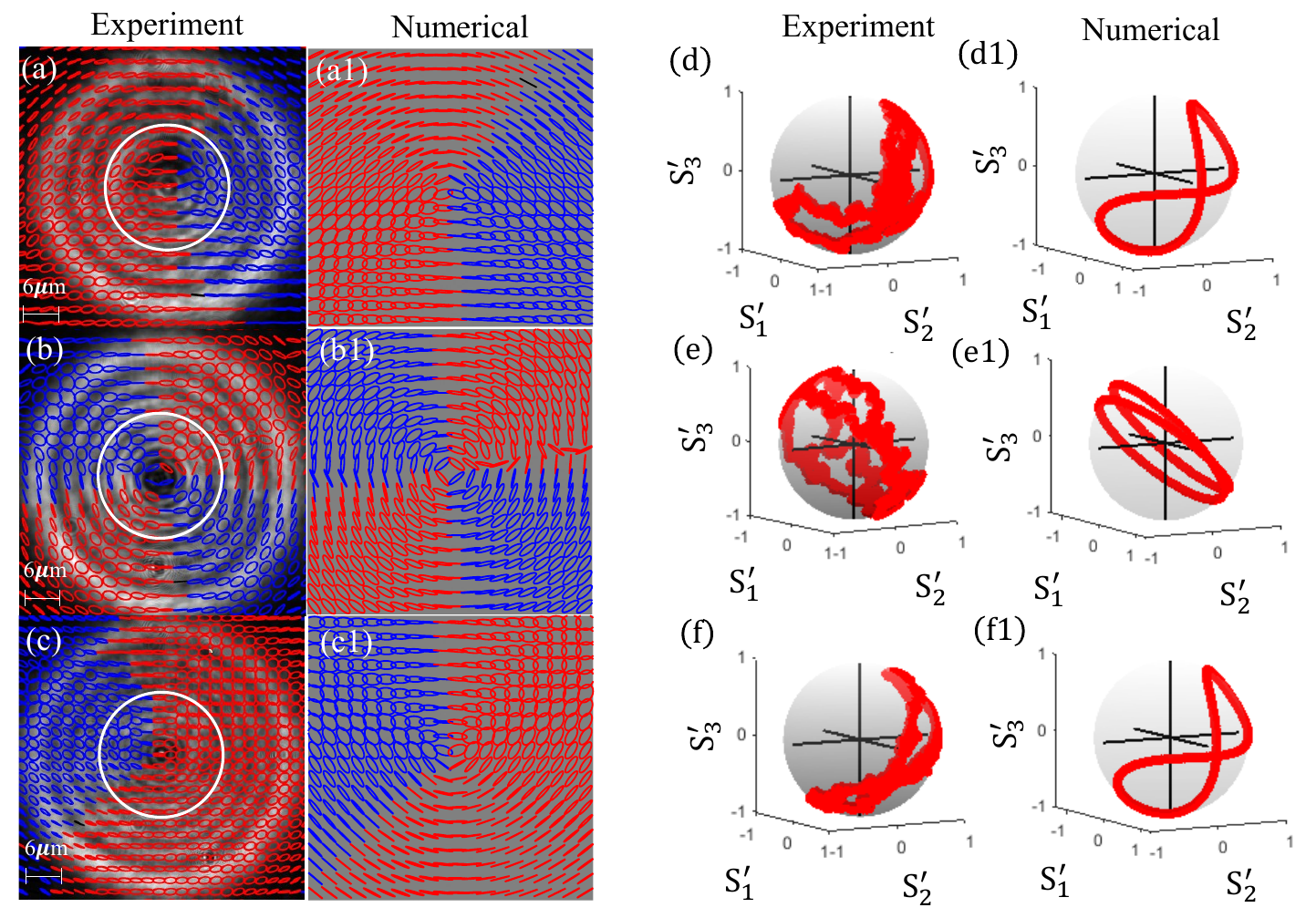}
\caption{\label{A1} (a–c) Experimental measurements at applied voltages of 1.8$V$, 2.2$V$, and 3.8$V$, respectively, shown together with numerical results (a1-c1) for retardations of $\delta \approx 1.5\pi, \pi$ and $\pi$. (d–f) Stokes parameters mapped onto the Poincar\'{e} sphere for the corresponding voltages in experiment and the respective $\delta$ values in simulation in (d1-f1).}
\end{figure}
Fig. \ref{A1} shows the experimental and numerical results of the $|V\rangle$ polarization Stokes parameter evolution on the fig:poincaresphere along the beam. We can see the \rotatebox[origin=c]{45}{$\infty$} loop in the cases of 1.8$V$ and 3.8$V$, just as we did in the $|H\rangle$ input example.

\begin{figure}[ht!]
\centering
\includegraphics[width=0.7\linewidth]{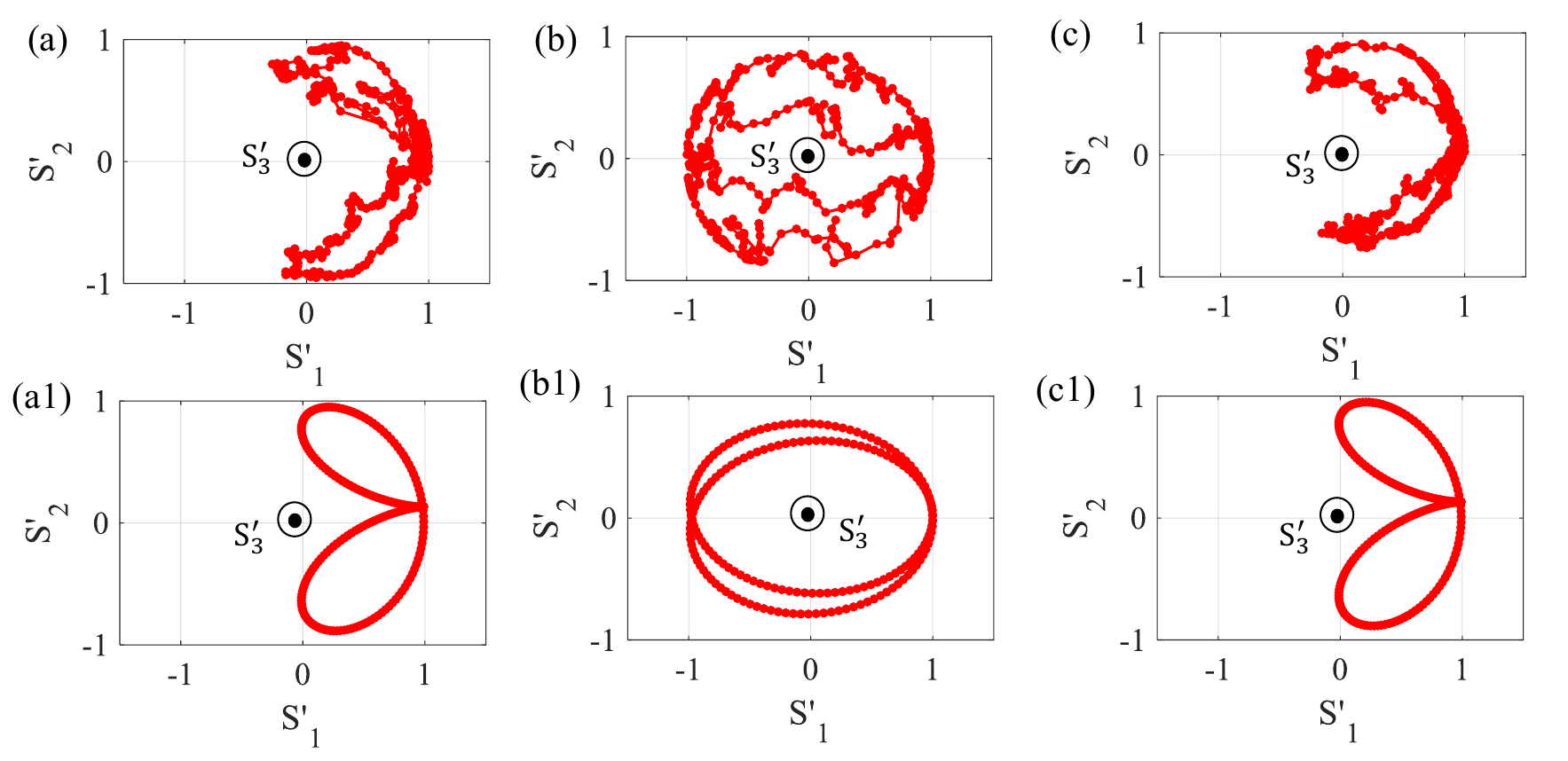}
\caption{\label{A2} Stokes parameters plotted on the $S'_2$–$S'_1$ plane through the $S'_3$ axis. Experimental results are shown for applied voltages $V_Q = 1.8V$, $2.2V$, and $3.8V$, together with numerical results corresponding to retardation $\delta \approx 1.5\pi$, $\pi$, and $0.5\pi$, respectively.}
\end{figure}

Fig. \ref{A2} shows the experimental and numerical evolution of the Stokes parameters on the fig:poincaresphere for a $|V\rangle$-polarized input. At applied voltages of $1.8V$ and $3.8V$, a \rotatebox[origin=c]{45}{$\infty$}-shaped trajectory is obtained, in which neither loop encloses the $S'_3$ axis. In contrast, at $2.2V$, both loops encircle the $S'_3$ axis.

\section*{A2: Stokes parameters for $|D\rangle$ $\&$ $|A\rangle$ input}

\begin{figure}[ht!]
\centering
\includegraphics[width=0.6\linewidth]{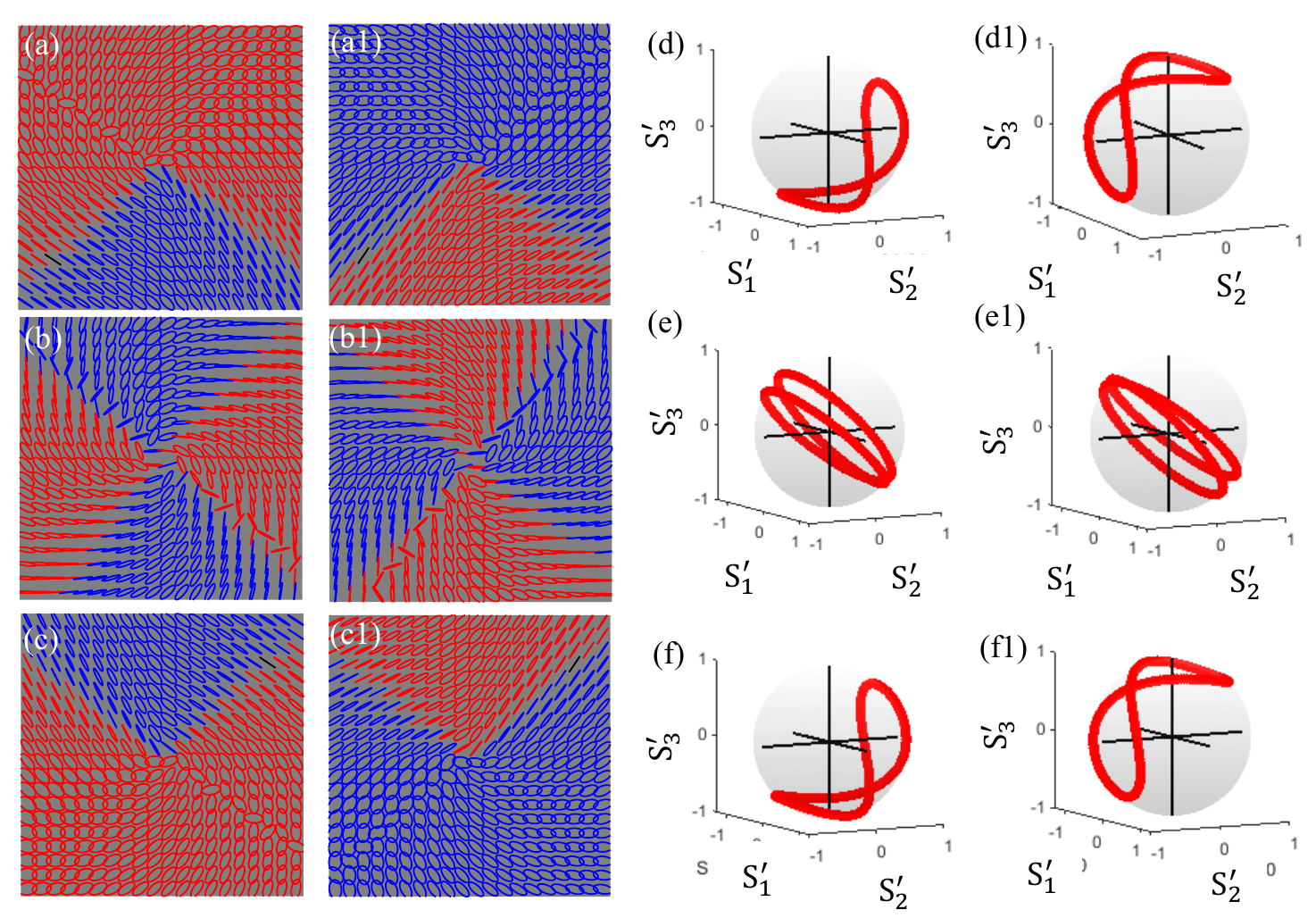}
\caption{\label{A3} Numerical Stokes-parameter evolution on the Poincar\'{e} sphere for (a–c) $|D\rangle$ and (a1–c1) $|A\rangle$ inputs. corresponding trajectories shown in (d–f) and (d1–f1), respectively.}
\end{figure}

Fig. \ref{A3} shows the numerical results of the $|D\rangle$ and $|A\rangle$ polarization Stokes parameter evolution on the fig:poincaresphere introduce by the $Q$-plate. We can see one of the loop of\rotatebox[origin=c]{45}{$\infty$} in the cases of $\delta \approx 1.5\pi$ and $\delta \approx 0.5\pi$ now encircling the $S'_3$ axis, while $\delta \approx \pi$ twice encirlce the $S'_3$.

\begin{figure}[ht!]
\centering
\includegraphics[width=0.5\linewidth]{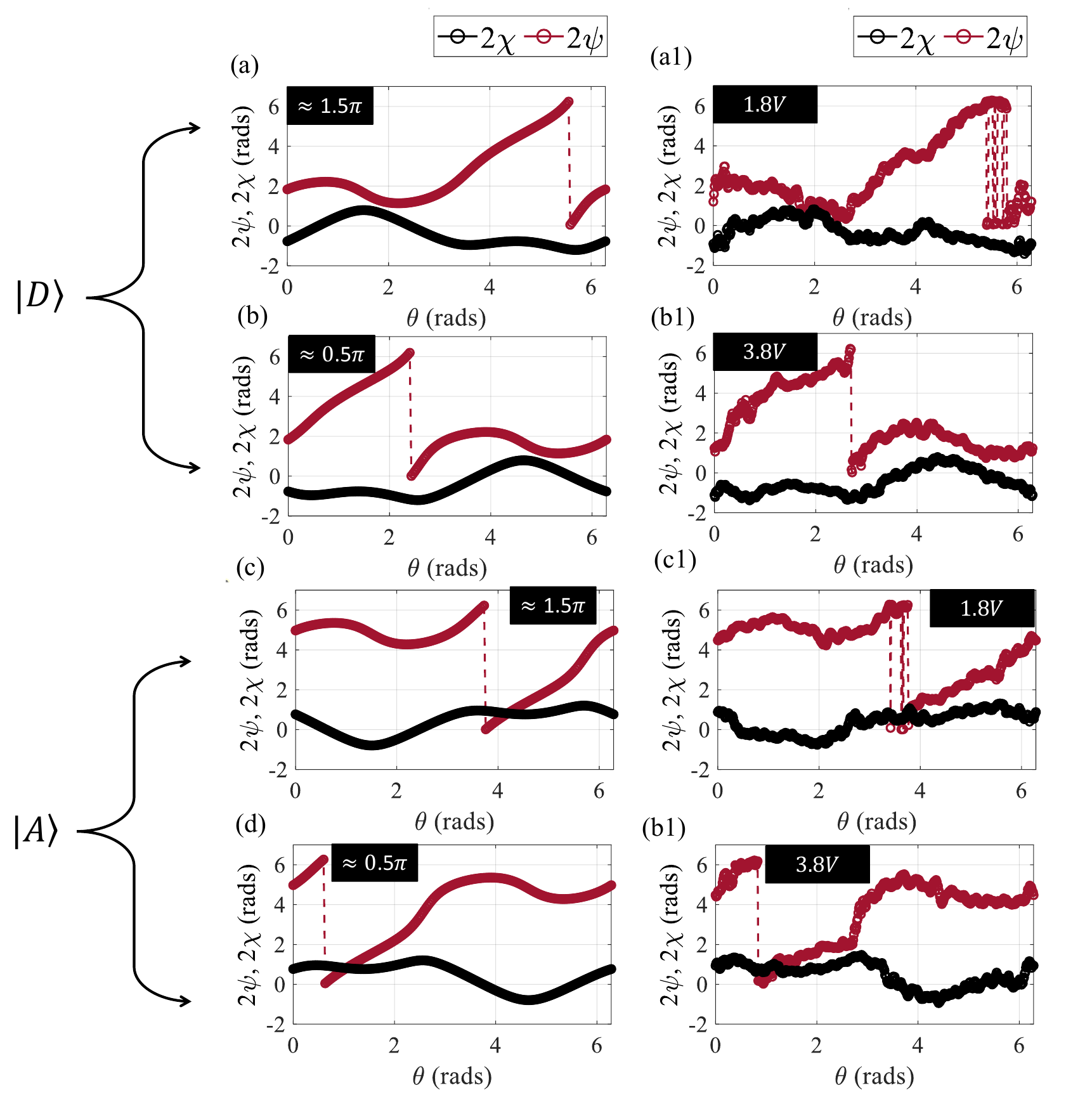}
\caption{\label{A4} (a,b) Numerical $2\chi$ and $2\psi$ vs. $\theta$ for $\delta \approx 1.5\pi$ and $0.5\pi$. (a1,b1) corresponding experimental results at 1.8 V and 3.8 V for $|D\rangle$. (c,d) Numerical results. (c1,d1) experimental results for $|A\rangle$ under the same conditions.}
\end{figure}

In fig .\ref{A4} shows the angular evolution of $2\psi$ and $2\chi$ around the beam. Panels (a,b) present numerical results and (a1,b1) the corresponding experimental results for the $|D\rangle$ input at $\delta \approx 1.5\pi$ and $0.5\pi$, corresponding to applied voltages of 1.8 V and 3.8 V, respectively. Panels (c,d) and (c1,d1) show the numerical and experimental results, respectively, for the $|A\rangle$ input under the same conditions.

\begin{figure*}[ht!]
\centering
\includegraphics[width=0.8\linewidth]{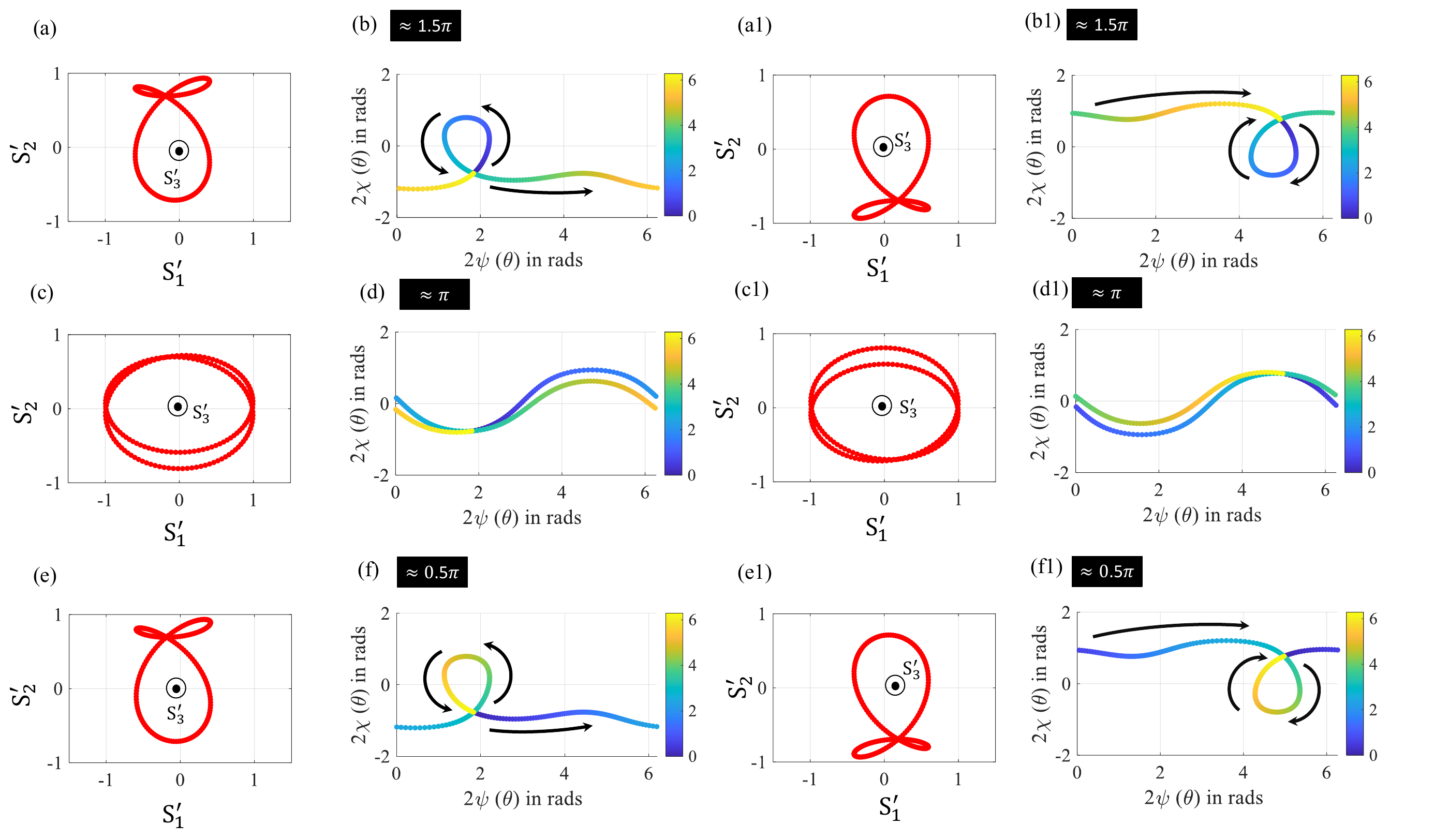}
\caption{\label{A5} Numerical results for Stokes-parameter trajectories on the $S'_2$–$S'_1$ plane and the corresponding $2\chi$–$2\psi$ plots for $|D\rangle$ and $|A\rangle$ inputs at $\delta \approx 1.5\pi$ (a–b, a1–b1), $\delta \approx \pi$ (c–d, c1–d1), and $\delta \approx 0.5\pi$ (e–f, e1–f1).}
\end{figure*}

Fig. \ref{A5} summarizes the numerical Stokes-parameter dynamics on the $S'_2$–$S'_1$ plane and the corresponding $2\chi$–$2\psi$ representations for $|D\rangle$ and $|A\rangle$ inputs. For $\delta \approx 1.5\pi$ and $0.5\pi$, the trajectories form a single loop encircling the $S'_3$ axis, leading to open single-loop structures in the $2\chi$–$2\psi$ plane. In contrast, at $\delta \approx \pi$, both loops encircle the $S'_3$ axis, and the corresponding $2\chi$–$2\psi$ plots exhibit oscillatory polarization dynamics with a sinusoidal form. The same qualitative behavior is observed for both input polarizations.

\bibliography{apssamp}

\end{document}